\documentclass[12pt,draftclsnofoot, onecolumn]{IEEEtran}		

\usepackage{times}
\usepackage{amsmath,dsfont,bm}
\usepackage{amssymb,amsthm}
\usepackage{epsfig,verbatim}
\usepackage{setspace}
\usepackage{color}
\usepackage{cite}
\usepackage{epstopdf}
\usepackage{acronym}
\usepackage[ruled,vlined,linesnumbered,lined,commentsnumbered]{algorithm2e}

\newtheorem*{theorem*}{Theorem}

\definecolor{NewColor}{rgb}{0,0,0}

\newcommand{\myVec}[1]{{\boldsymbol{#1}}}

\newcommand{\mySet}[1]{\mathcal{#1}}

\newcommand{\XT}{X_T}
\newcommand{\XP}{X_{\phi}}
\newcommand{\nodeidx}{j}
\newcommand{\thetavec}{\bm{\theta}}
\newcommand{\psip}{\psi_{\phi,i}}
\newcommand{\thetap}{\thetavec_{\phi,i}}
\newcommand{\psif}{\psi_{T,i}}
\newcommand{\thetaf}{\thetavec_{T,i}}
\newcommand{\epsp}{\varepsilon_{\phi}}
\newcommand{\epsT}{\varepsilon_{T}}

\newcommand{\psiptheta}{\psip^{(\thetap)}}
\newcommand{\psiftheta}{\psif^{(\thetaf)}}

\newcommand{\dsone}{\mathds{1}}

\newcommand{\dsU}{\mathds{U}}

\newcommand{\mN}{\mathcal{N}}
\newcommand{\mI}{\mathcal{I}}
\newcommand{\mZ}{\mathcal{Z}}

\newcommand{\mX}{\mathcal{X}}
\newcommand{\mD}{\mathcal{D}}

\newcommand{\bphi}{\bar{\phi}}
\newcommand{\bT}{\bar{T}}

\acrodef{pc}[PC]{pulse-coupling}
\acrodef{mac}[MAC]{medium access control}
\acrodef{ml}[ML]{machine learning}
\acrodef{std}[STD]{standard deviation}
\acrodef{los}[LOS]{line-of-sight}
\acrodef{gps}[GPS]{Global Positioning System}
\acrodef{nn}[NN]{neural network}
\acrodef{dnn}[DNN]{deep neural network}
\acrodef{wsn}[WSN]{wireless sensor network}
\acrodef{hd}[HD]{half-duplex}
\acrodef{fd}[FD]{full-duplex}
\acrodef{rv}[RV]{random variable}
\acrodef{dt}[DT]{discrete-time} 
\acrodef{ct}[CT]{continuous-time} 
\acrodef{wscs}[WSCS]{wide-sense cyclostationary}
\acrodef{wsacs}[WSACS]{wide-sense almost cyclostationary}
\acrodef{ofdm}[OFDM]{orthogonal frequency division multiplexing}
\acrodef{tdma}[TDMA]{time division multiple access}
\acrodef{cdma}[CDMA]{code division multiple access}
\acrodef{noma}[NOMA]{non-orthogonal multiple access}
\acrodef{asmcgn}[AS-MCGNC]{asynchronously-sampled memoryless cyclostationary Gaussian noise channel}
\acrodef{cmt}[CMT]{continuous mapping theorem}
\acrodef{iot}[IOT]{internet of things}
\acrodef{lti}[LTI]{linear, time-invariant}
\acrodef{iiot}[IIOT]{industrial internet of things}
\acrodef{ptp}[PTP]{Precision Time Protocol}
\acrodef{ntp}[NTP]{Network Time Protocol}
\acrodef{gptp}[gPTP]{generalized Precision Time Protocol}
\acrodef{ptcp}[PTCP]{Precision Transparent Clock Protocol}
\acrodef{ptcpkf}[PTCP-KF]{Precision Transparent Clock Protocol with Kalman filter}
\acrodef{dl}[DL]{deep learning}
\acrodef{pco}[PCO]{pulse-coupled oscillator}
\acrodef{ftsp}[FTSP]{flooding time synchronization protocol}
\acrodef{pll}[PLL]{phase-locked loop}
\acrodef{drl}[DRL]{deep reinforcement learning}
\acrodef{vcc}[VCC]{voltage controlled clock}
\acrodef{td}[TD]{time difference detector}
\acrodef{npd}[NPD]{normalized phase difference}
\acrodef{pdd}[PDD]{phase difference detector}
\acrodef{gd}[GD]{gradient descent}
\acrodef{pd}[PD]{phase discriminator}
\acrodef{spd}[SPD]{signal power detector}
\acrodef{mlp}[MLP]{multi-layered perceptron}
\acrodef{dasa}[DASA]{\ac{dnn}-aided synchronization algorithm}
\acrodef{mbgd}[MB-SGD]{mini-batch stochastic gradient descent}
\acrodef{npdr}[NPDR]{\ac{npd} range}
\acrodef{pfdsa}[PFDSA]{phase and frequency \ac{dnn}-aided synchronization algorithm}

\newif\ifcomments
\definecolor{CmtColor}{rgb}{0,0.6,1}

\long\def\symbolfootnote[#1]#2{\begingroup\def\thefootnote{\fnsymbol{footnote}}\footnote[#1]{#2}\endgroup}

\IEEEoverridecommandlockouts
\IEEEaftertitletext{\vspace{-1\baselineskip}}

\title{Model-Based Learning for Network Clock Synchronization in Half-Duplex TDMA Networks 
}

\author{ 

	Itay Zino, Ron Dabora, \IEEEmembership{Senior Member, IEEE}, and H. Vincent Poor, \IEEEmembership{Fellow, IEEE}
	
	\thanks{  I. Zino and R. Dabora  are with the School of Electrical and Computer Engineering, Ben-Gurion University of the Negev, Be'er-Sheva, Israel (e-mail:  itayzin@post.bgu.ac.il;  daborona@bgu.ac.il); H. V. Poor (email: poor@princeton.edu) is with the Department of Electrical and Computer Engineering, Princeton University, Princeton, NJ, USA. R. Dabora is currently a Visiting Fellow at Princeton University. 
 This work was supported in part by the Israel Science Foundation under Grant 584/20, by the Israel Ministry of Economy via the 5G-WIN Consortium, and by the U.S National Science Foundation under Grants CNS-2128448 and ECCS-2335876.  © © 20xx IEEE. Personal use of this material is permitted. Permission from IEEE must be obtained for all other uses, in any current or future media, including reprinting/republishing this material for advertising or promotional purposes, creating new collective works, for resale or redistribution to servers or lists, or reuse of any copyrighted component of this work in other works. }

}

\begin{document}
	
	\maketitle
	\thispagestyle{empty}

\begin{abstract}
    \color{black}
     Supporting increasingly higher rates in wireless networks requires highly accurate clock synchronization across the nodes. Motivated by this need,  in this work we consider distributed clock synchronization for half-duplex (\acs{hd}) \acs{tdma} wireless networks. We focus on pulse-coupling (PC)-based synchronization as it is practically advantageous for high-speed networks using low-power nodes. Previous works on PC-based synchronization for \acs{tdma} networks assumed full-duplex communications, and focused on correcting the clock phase at each node, without synchronizing clocks' frequencies. However, as in the  \acs{hd} regime corrections are temporally sparse, uncompensated clock frequency differences between the nodes result in large phase drifts between updates. 
     Moreover, as the clocks determine the processing rates at the nodes, leaving the clocks' frequencies unsynchronized results in processing rates mismatch between the nodes, leading to a throughput reduction. Our goal in this work is to synchronize both clock frequency and clock phase across the clocks  in \acs{hd} \acs{tdma} networks, via distributed processing. The key challenges are the coupling between frequency correction and phase correction, and the lack of a computationally efficient analytical framework for determining the optimal correction signal at the nodes. We address these challenges via a \ac{dnn}-aided nested loop structure in which the \acp{dnn} are used for generating the weights applied to the loop input for computing the correction signal. This loop is operated in a sequential manner which decouples frequency and phase compensations, thereby facilitating synchronization of both parameters. Performance evaluation shows that the proposed scheme significantly improves synchronization accuracy compared to the conventional approaches.
	
\end{abstract}
	
\acresetall
	

	\section{Introduction}
	\label{sec:Intro}

Clock synchronization is critical for the  implementation of high-rate communications over wireless networks operating in a \ac{tdma} regime, as higher throughput requires stricter temporal coordination between the nodes.
The common approach for clock synchronization in ad-hoc wireless networks is based on exchanging packets carrying timing information. While this approach is relatively simple to implement, it is susceptible to random queuing delays at the \ac{mac} layer as well as to random processing delays at the nodes, which limits its accuracy. Moreover, the processing associated with packet-based synchronization entails a higher power consumption at the nodes~\cite{simeone2008distributed}. 

An alternative approach for network clock synchronization is based on \ac{pc}, in which the nodes transmit signature sequences, and each node applies processing based on the time stamps it assigns to the signatures received from the other nodes. Commonly, a clock update signal is computed by weighting the received time stamps  according to their relative received power w.r.t. to the received powers from the other nodes, see, e.g., \cite{simeone2008distributed}. Such weighting leads to good performance when all clocks have the same frequency as long as the propagation delays between the nodes are negligible. The major challenges thus follow as the accuracy of \ac{pc}-based algorithms is strongly dependent on the weights, yet the optimal weights for practical scenarios, in which there are frequency differences between the clocks as well as propagation delays, are unknown. This has motivated our work  \cite{ShlezingerDabora:2023}, which proposed a \ac{dnn}-aided synchronization scheme in which the \ac{dnn} is trained to compute  the weights at the nodes. The algorithm, presented in \cite{ShlezingerDabora:2023}, improves the accuracy by two orders of magnitudes compared to the classic weights of~\cite{simeone2008distributed}.

While previous works,  \cite{simeone2008distributed}, \cite{ShlezingerDabora:2023}, \cite{simeone2007distributed}, have proposed distributed algorithms which achieve different clock synchronization accuracy across the nodes, they all share a common limitation: In all  these works, the correction signal is applied to the clock's phase, {\em without synchronizing the clock's frequency}.  The main benefit of this approach is  simplicity as only the clock phase is updated, and when applied in networks operating in the \ac{fd} regime, it can indeed attain good synchronization performance.  
However, when applied in the common scenario of \ac{hd} communications,  rendering the frequencies of the clocks at the nodes unsynchronized results in significant drifts in the clock phase between updates across the nodes in the network, which decreases the throughput of the network.
As the clocks drive the processing at the nodes, then, when the clocks' frequencies remain unsynchronized, different nodes carry out processing at different rates, 
which will lead to loss of information even with perfect clock phase synchronization between the nodes. In order to mitigate these issues, {\em clock frequency synchronization has to be achieved along with clock phase synchronization}  
in \ac{hd} \ac{tdma} networks, which is the focus of this work. We note that such network architectures are relevant to multiple network and application scenarios, including aerial networks \cite{AerialNetworks:2020}, cooperative navigation in wireless networks \cite{TDMAnavigation:2020}, and sensor networks \cite{TDMASenser:2020}.

\textit{\textbf{Main Contributions:}} In this work, we propose a new \ac{pc}-based \ac{dnn}-aided synchronization algorithm that  synchronizes both the frequencies and the phases of the clocks, which makes it particularly suitable  for \ac{hd} \ac{tdma} wireless networks. 
As coupling between phase correction and frequency correction may result in  frequency differences also after convergence,  we implement a staggered update rule which decouples the phase updates from the frequency updates. We then propose {\em an unsupervised} and {\em distributed} training scheme for obtaining the weights. The main novelty here is facilitating the training of both the frequency and the phase detectors overcoming the above mentioned coupling. The proposed \ac{dnn}-aided algorithm implements a model-based learning approach, which benefits from the stability and convergence properties of the loop architecture, see, e.g., \cite{ShlezingerDabora:2023}. As training is carried out online with the actual deployment, the learned \ac{dnn} coefficients 
  result in superior synchronization performance compared to other distributed approaches. 
The proposed scheme is compared with a direct extension of the classic algorithm \cite{simeone2008distributed} to \ac{hd} operation and frequency synchronization, as well as with an adapted version of \cite{maggs2012consensus}. The comparison shows that the synchronization accuracy of the \ac{dnn}-aided scheme is considerably higher than that of analytical scheme \cite{simeone2007distributed}, while the adapted scheme~of \cite{maggs2012consensus} is not able to synchronize either the frequency or the phase.

 \textit{\textbf{Organization}}: The rest of this work is organized as follows: In Section \ref{sec:Clk_model}  the clock model and the baseline network model are described  and the classic algorithm \cite{simeone2007distributed} is extended to operate in the \ac{hd} regime and to achieve frequency synchronization. 
 The performance of the extended scheme is then demonstrated to motivate the current work; 
 Section \ref{sec:Algorithm} details the proposed algorithm and Section \ref{sec:Training Scheme} details the associated training scheme; Section \ref{sec:simulations} presents simulation tests and discussion; Lastly, Section \ref{sec:conclusion} concludes the work.



\section{Preliminaries: Distributed Pulse-Coupled Time Synchronization for Wireless Networks}
\label{sec:Clk_model}

\subsection{A Baseline Scenario}
\label{subsec:baseline}
\subsubsection{Network and Clock Models}

We consider a network with $N$ nodes, indexed by $i\in\{1, 2, ..., N\}\triangleq\mI_N$. Each node $i$ has a clock oscillator with its own inherent period, denoted by $T_i$. Ignoring phase noise (see, e.g., \cite[Sec. V]{maggs2012consensus}, \cite{simeone2008distributed}), then, at update index $k\in\mZ$ the  clock time at node $i$, $\phi_i[k]$,  can be expressed~as 
	\begin{equation}
            \phi_i[k]=\phi_i[0]+k\cdot T_i,
	\end{equation}
where $\phi_i[0]$ denotes the clock time at index $k=0$.

\subsubsection{Scenario Parameters}
\begin{figure}
		\centering
		\includegraphics[width=0.82\linewidth]{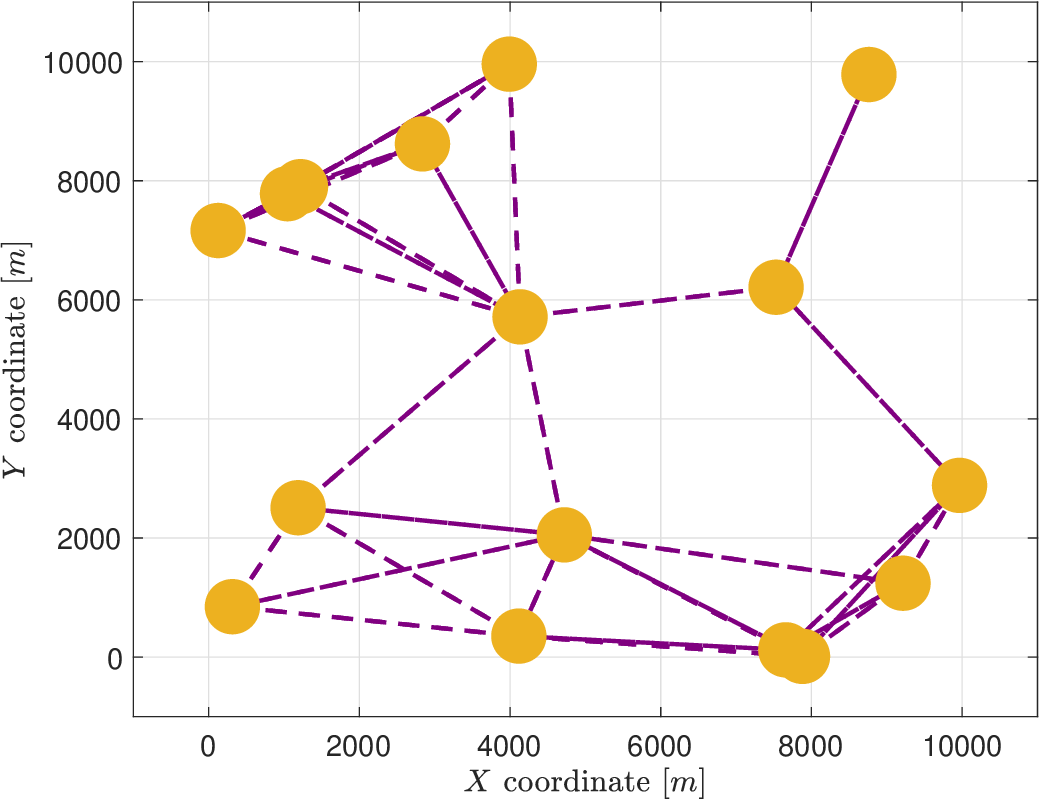}
        \caption{\small Nodes placement for the baseline scenario (nodes are denoted by circles). Links over which the signal is received at the respective destination above the reception threshold are denoted by lines.}
        \label{fig:Geographical_Locations}
    \end{figure}

 As a baseline example, we set the network size to $N=16$ nodes, deployed in a square area with a side-length of $10\mbox{ [Km]}$.  We assume $2$-ray propagation with antenna height $1.5$ [m], transmit power of $33$ [dBm], and a reception threshold of $P_{th}=-114$ [dBm]. With the node deployment depicted by the circles in Fig. \ref{fig:Geographical_Locations}, it follows that $30\%$ of the links carry signals that are received at their respective destinations above the reception threshold. These links are marked by the lines in Fig. \ref{fig:Geographical_Locations}. The nominal value of the \ac{tdma} period is set to $5$ [msec] and the respective nominal \ac{tdma} frequency is $f_{\rm nom} = 200 \; [\mbox{Hz}]$. We model clock uncertainty by generating the initial \ac{tdma} frequency $f_i$ at each node $i$  randomly, according to a uniform distribution $f_i \sim \dsU[f_{\rm nom}(1 - 150 \cdot 10^{-6}), f_{\rm nom}(1 + 150 \cdot 10^{-6})]$, which corresponds to clock accuracy of 150 [ppm]. The initial \ac{tdma} period at node $i$ is obtained as  $T_i = \frac{1}{f_i}$.  Each node is then assigned an initial phase $\phi_i[0]$ uniformly distributed over $[0,T_i]$. After startup, the nodes begin to transmit their signatures, such that at time $k$, node $i=(k\!\! \mod{\!N})+1$ transmits and the remaining $N-1$ nodes receive. Each receiving node assigns a time stamp and a received signal power value to each received signature. At time indices $k$ such that $(k\!\! \mod{\!N})\! =\! N-1$, each node independently processes its received time stamps and received signal power values to generate correction signals applied to update its \ac{tdma} clock period and clock phase. Accordingly, in the following we let $T_i[k]$ denote the period of the clock at node $i$ at time index~$k$, and $T_i[0]$ denote the initial period value.

\subsection{Phase and Frequency Synchronization with Current State-of-the-Art}
\label{subsec:stateoftheart}

\begin{figure}
		\centering
		\includegraphics[width=0.90\linewidth]{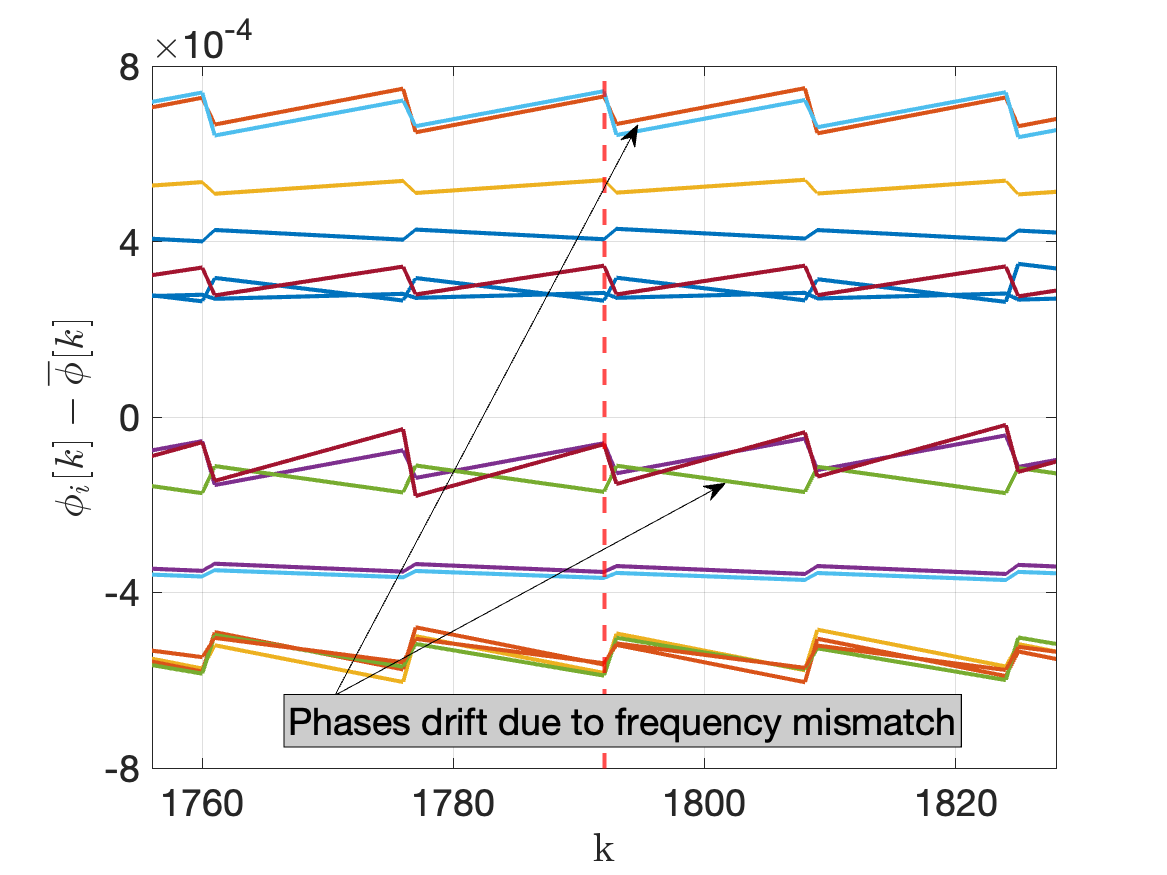}
        \caption{\small Clock phases in an HD-TDMA network without period update. Each line depicts the clock phase at one node.}
		\label{fig:half_duplex_no_period_correction}
    \end{figure}

As mentioned in Sec. \ref{sec:Intro}, the classic  scheme of \cite{simeone2008distributed} operates to achieve network clock synchronization by updating only the clock phase leaving the frequencies of the clocks at the nodes unsynchronized. 
Fig. \ref{fig:half_duplex_no_period_correction} depicts the clock phases obtained with the classic algorithm \cite{simeone2008distributed} in the \ac{hd} regime, where the phase correction is applied at times $k$ s.t. $(k \mod{N}) = N-1$. The phase is plotted as the offset w.r.t. to the mean instantaneous phase, defined as $\bphi[k]\triangleq \frac{1}{N}\sum_{i=1}^N\phi_i[k]$. 
We observe that the lack of frequency synchronization induces  significant phase drifts across the nodes. We measure the network synchronization accuracy via the \ac{npdr}, 
which is the maximal clock phase difference between any pair of clocks in the network, normalized to the instantaneous mean period, defined as  $\bT[k]\triangleq \frac{1}{N}\sum_{i=1}^N T_i[k]$: 
\begin{eqnarray}
	    \mbox{NPDR}[k] & \triangleq &  \Big(\mathop{\max}\limits _{i_1\in \mI_N} \phi_{i_1}[k] - \mathop{\min}\limits _{i_2\in \mI_N} \phi_{i_2}[k]\Big)/\bT[k].
	    \label{eqn:NPDrange}
\end{eqnarray}
A straightforward approach for achieving clock phase and frequency synchronization in \ac{hd} networks is to generalize the model of \cite{simeone2008distributed} by adding a frequency synchronization loop, using the same power-dependent weights. Before stating the generalization we  recall the clock phase update rule at node $i$, as stated in \cite[Eqn. (16)]{simeone2008distributed}:
\begin{equation}
    \label{eqn:Simeone_Original}
    \phi_i[k+1] = \phi_i[k] + T_i[0] + \varepsilon_0\!\! \underset{j \in \mathcal{N}(i)}{\sum} \!\!\alpha_{i,j} \cdot (\phi_j[k] + q_{i,j} - \phi_i[k]),
\end{equation}
where $\varepsilon_0$ denotes the loop gain, $q_{i,j}$ is the propagation delay between nodes $i$ and $j$, and $\mN(i)$ is the neighborhood set of node $i$, consisting of all the nodes whose signals are received at node $i$ with power higher than $P_{th}$. Note that here the clock period is fixed at $T_i[0]$ as this value is not updated by the algorithm \cite{simeone2008distributed}.
Let $ P_{i,j}[k]$ denote the power of the signal received at node $i$ from node $j$ at time index $k$, and denote by 
$t_{i,j}[k]\triangleq \phi_j[k] + q_{i,j}$ the time stamp node $i$ assigns to the signature received from node $j$.  At startup, note $i$ initializes the storage variables $\Delta\phi_{i,j}^{\rm prev.} =  \Delta\phi_{i,j} = P^{(i)}[j] = 0$.
At time index $k$, node $j = (k \mod{N}) + 1$ transmits and nodes $i\in\mI_N \setminus j$ receive. 
Then, for any node $i\in\mI_N \setminus j$ for which $P_{i,j}[k]>P_{th}$, the node first updates  $\mN(i)$ to include $j$, and then computes and stores $\Delta\phi_{i,j}^{\rm prev.}\leftarrow\Delta\phi_{i,j}$,  $\Delta\phi_{i,j} \leftarrow t_{i,j}[k]-\phi_i[k]$, and $P^{(i)}[j] \leftarrow P_{i,j}[k]$. At times $k$ s.t. $(k\!\! \mod{\!3N}) = 2N-1$, the weights at node $i$ are set to \cite[Eqn. (8)]{simeone2008distributed}
\[
    \alpha_{i,j} = P^{(i)}[j] / \underset{m \in \mathcal{N}(i)}{\sum}P^{(i)}[m].
\]

\begin{figure}
		\centering
		\includegraphics[width=0.9\linewidth]{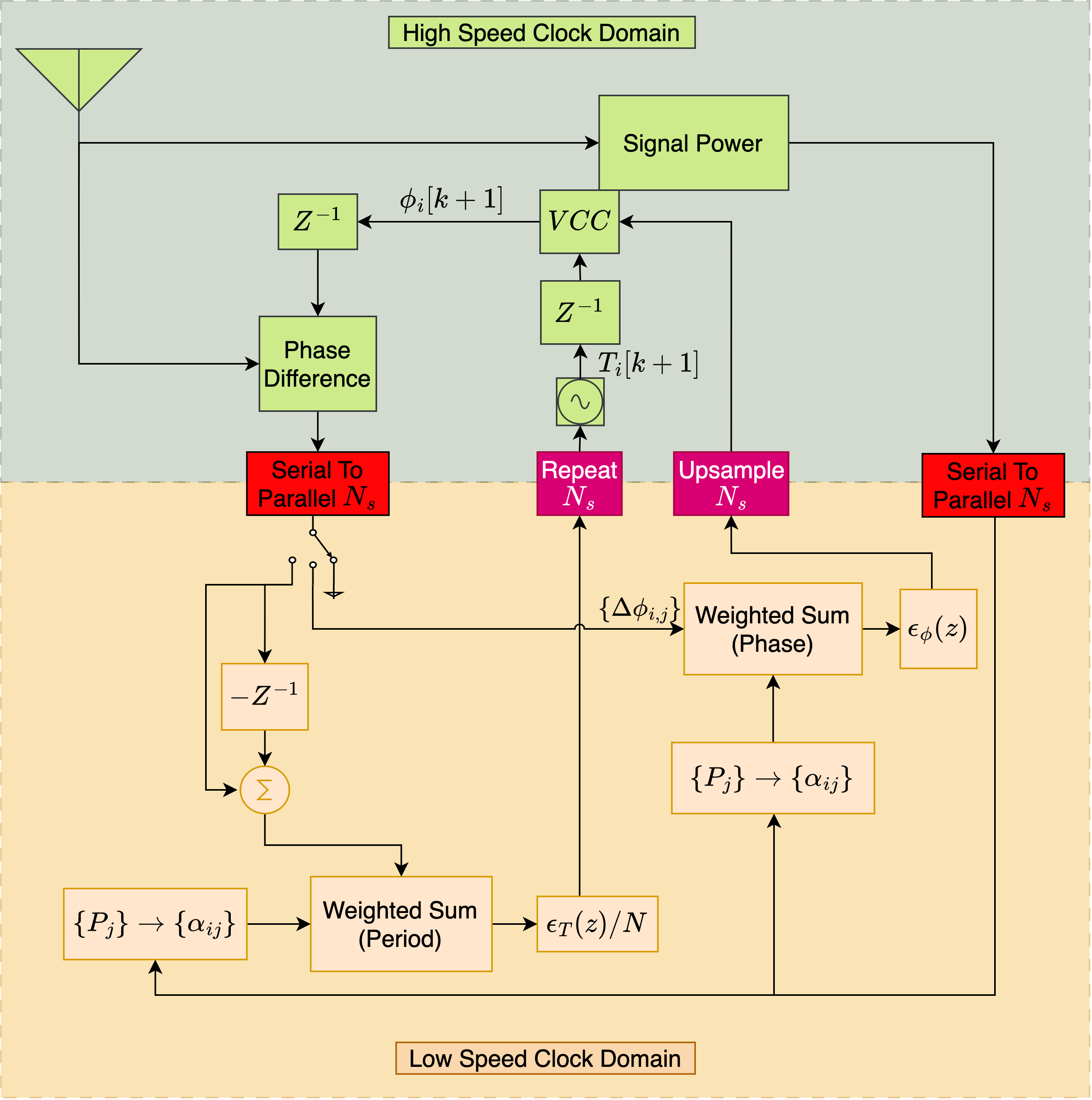}
     		\caption{\small Schematic description of the ESSBS algorithm (Eqns. \eqref{eqn:Simeone_Extended_ph}--\eqref{eqn:Simeone_Extended_T}).} 
		\label{fig:DT_classic_PLL}
    \end{figure}

\begin{figure}
		\centering
		\includegraphics[width=0.98\linewidth]{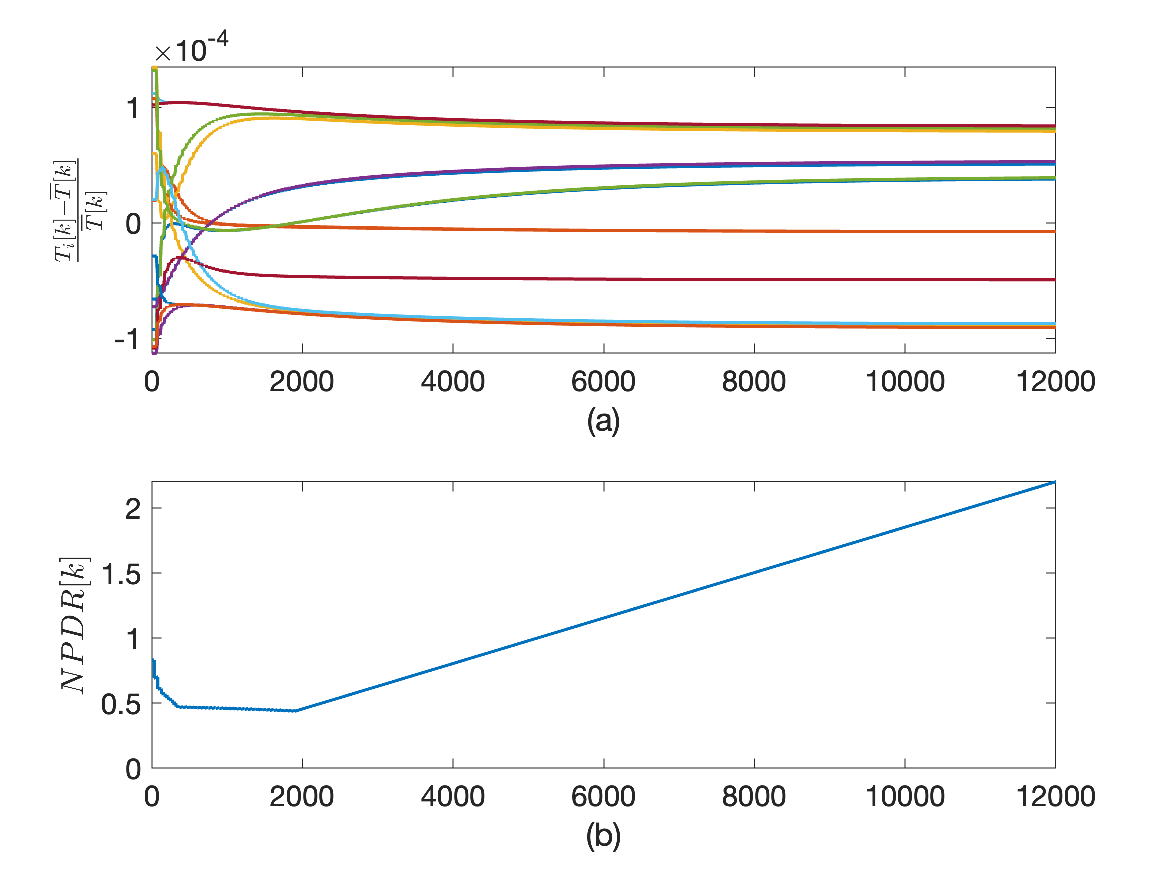}
 		\caption{\small Performance of the ESSBS algorithm: (a) The evolution of the clock periods $T_i[k]$ vs. $k$, where each line represents the evolution of the clock period at a different node in the network; (b) The NPDR of the network vs. $k$.} 
		\label{fig:Test_ESSBS}
    \end{figure}

In the following we  generalize the algorithm of \cite{simeone2008distributed} to include period synchronization
by introducing a nested loop designed to synchronize $T_i[k]$ across the different nodes. The generalized scheme, referred to in this work as the extended Simeone-Spagnolini-BarNess-Strogatz (ESSBS) algorithm, operates as follows: 
\begin{align}
    \label{eqn:Simeone_Extended_ph}
    &\!\!\!\!\phi_i[k+1] = \phi_i[k] + T_i[k] + Q_{\phi,i}[k]\notag\\
    & \!\!\!\! Q_{\phi,i}[k]=\left\{\begin{array}{cl} 0 & ,(k \!\!\!\!\mod{3N}) \ne 3N-1\\
    \!\!\epsp\!\!\!\! \underset{m \in \mathcal{N}(i)}{\sum}\!\!\!\! \alpha_{i,m} \cdot \Delta\phi_{i,m} &  ,(k\!\!\!\! \mod{3N}) = 3N-1\end{array} \right.\\
     \label{eqn:Simeone_Extended_T}
     &\!\!\! T_i[k+1]  = T_i[k] + Q_{T,i}[k]/N\notag\\
     & \!\!\!\! Q_{T,i}[k]\!=\!\!\left\{\begin{array}{cl} \!\!\!\!\!\!0 & \!\!\!,0\le (k \!\!\!\!\mod{3N}) < 2N\!\!-\!\!1\\
    \!\!\epsT \!\!\!\!\underset{m \in \mathcal{N}(i)}{\sum} \!\!\!\!\!\alpha_{i,m}(\Delta\phi_{i,m} & \\
    \;\;\;\;\;\;\;\!\!\!\! - \Delta\phi_{i,m
    }^{\rm prev.}) &  \!\!\!,(k\!\!\!\! \mod{3N}) = 2N-1\\
      \!\!\!\!\!\!\!\!\!\!\!Q_{T,i}[k-1] &   \!\!\!,2N\!\le\!(k\!\!\!\! \mod{3N})\! <\!3N\!\!-\!\!1
    \end{array} \right. \!
\end{align}
Fig. \ref{fig:DT_classic_PLL} presents a schematic depiction of the ESSBS algorithm.
In Sec. \ref{sec:Algorithm} we elaborate on the factor of $3$ included in the  modulo operations in \eqref{eqn:Simeone_Extended_ph} and  \eqref{eqn:Simeone_Extended_T}.
Fig. \ref{fig:Test_ESSBS} depicts the clock periods and the \ac{npdr} achieved by the ESSBS algorithm for the baseline scenario detailed in Sec. \ref{subsec:baseline}. Observe that ESSBS is not able to achieve accurate clock synchronization, and in fact, as time increases the clocks' phases  drift farther apart. 
{\em This motivates the introduction of a \ac{dnn}-based weight computations}, considered in the next section.

\section{A Distributed \texorpdfstring{\ac{dnn}}{DNN}-Aided Frequency and Phase   Synchronization Scheme for the  HD Regime}
\label{sec:Algorithm}
We propose to improve upon the current state-of-the-art \ac{fd} algorithms \cite{simeone2008distributed} (which uses weights computed analytically) and \cite{ShlezingerDabora:2023} (which uses weights generated via a \ac{dnn}) by incorporating three new elements: (1) A \ac{dnn}-aided {\em frequency synchronization loop}  which operates together with the phase synchronization loop; (2) Modification of the  operation of the overall synchronization scheme, adapting it to the {\em \ac{hd} regime}; and (3) Introduction of an {\em unsupervised} and {\em distributed} training scheme that facilitates training of both \acp{dnn} without requiring the exchange of feedback information between the nodes. Next, we elaborate on the operation of the proposed combined frequency  and phase synchronization scheme in the \ac{hd} regime.

By analyzing (this original analysis, reported in \cite{Zino:Inpreparation}, is not included here due to space constraints)  the impact of propagation delays on the phase update rule in the ESSBS algorithm  \eqref{eqn:Simeone_Extended_ph} it was concluded that these delays induce  a drift in the frequency synchronization loop. Accordingly,  our proposed synchronization scheme, referred to as the {\em \ac{pfdsa}}, is designed to guarantee that  both the clock phases and periods are simultaneously synchronized.
We employ learning only of the weights, keeping the other elements of the loop structure 
of the ESSBS algorithm \eqref{eqn:Simeone_Extended_ph}-\eqref{eqn:Simeone_Extended_T} unchanged, see Fig. \ref{fig:DT_classic_PLL}.
This approach, known as a {\em model-based} approach, benefits from the desired convergence and stability properties associated with  the loop structure, while optimizing only the element whose analytical determination is suboptimal.

Let $\psif^{(\thetaf)}(\cdot)$ and $\psip^{(\thetap)}(\cdot)$ denote the \acp{dnn} used for computing the weights for the period and for the phase update loops at node $i\in\mI_N$, respectively,  where $\thetaf$ and $\thetap$ denote their respective \ac{dnn} parameters. In the \ac{hd} regime, a node requires {\em $N$ time slots} to collect the timing information from all of its neighbors (received above the reception threshold), which is a significantly slower rate than in the \ac{fd} regime, in which timing information from all received nodes is collected at {\em every} time slot. Accordingly, at time indices $k$ such that $(k \mod{N}) \ne N-1$,  node $i\in \mI_N$ updates its clock phase and frequency via 
\begin{subequations}
\label{eqn:Update_woCorrection}
    \begin{eqnarray}
    \label{eqn:phase_update}
        \phi_i[k+1] & = & \phi_i[k] + T_i\left[k\right] + \Omega_i[k]\\
        \label{eqn:period_update}
        T_i[k+1] & = & T_i[k] +A_i[k]/N,
    \end{eqnarray}
    
\end{subequations}
where $A_i[k]$ denotes the period correction term at node $i$ at time $k$, and $\Omega_i[k]$ is the corresponding phase correction term. In addition to updating their clocks, at time slot $k$  the node with index $j=(k \mod{N})+1$ transmits its synchronization signature  while the remaining $N-1$ nodes, $i\in\mI_N\setminus j$ are set to receive. 
Let $\XP^{(i)}[j]$ denote the phase difference between nodes  $j$ and $i$, stored at node $i$ at the {\em previous} update cycle.
Upon reception, each node $i\in\mI_N\setminus j$ for which $P_{i,j}[k]>P_{th}$,  stores $P^{(i)}[j]\leftarrow P_{i,j}[k]$, otherwise it sets $P^{(i)}[j]=0$. A node $i$ for which $P^{(i)}[j]>0$ computes and stores the period difference w.r.t. node $j$, defined as 
$\XT^{(i)}[j]\leftarrow \frac{1}{N}\big(t_{i,j}[k]-\phi_i[k] - \XP^{(i)}[j]\big)$. 
Node $i$ then proceeds to store  the new receive phase difference 
$\XP^{(i)}[j]\leftarrow t_{i,j}[k]-\phi_i[k]$.  
If   $P^{(i)}[j] = 0$, then node $i$ sets $\XP^{(i)}[j] = \XT^{(i)}[j]= 0$.

\begin{figure}
		\centering
		\includegraphics[width=0.9\columnwidth]{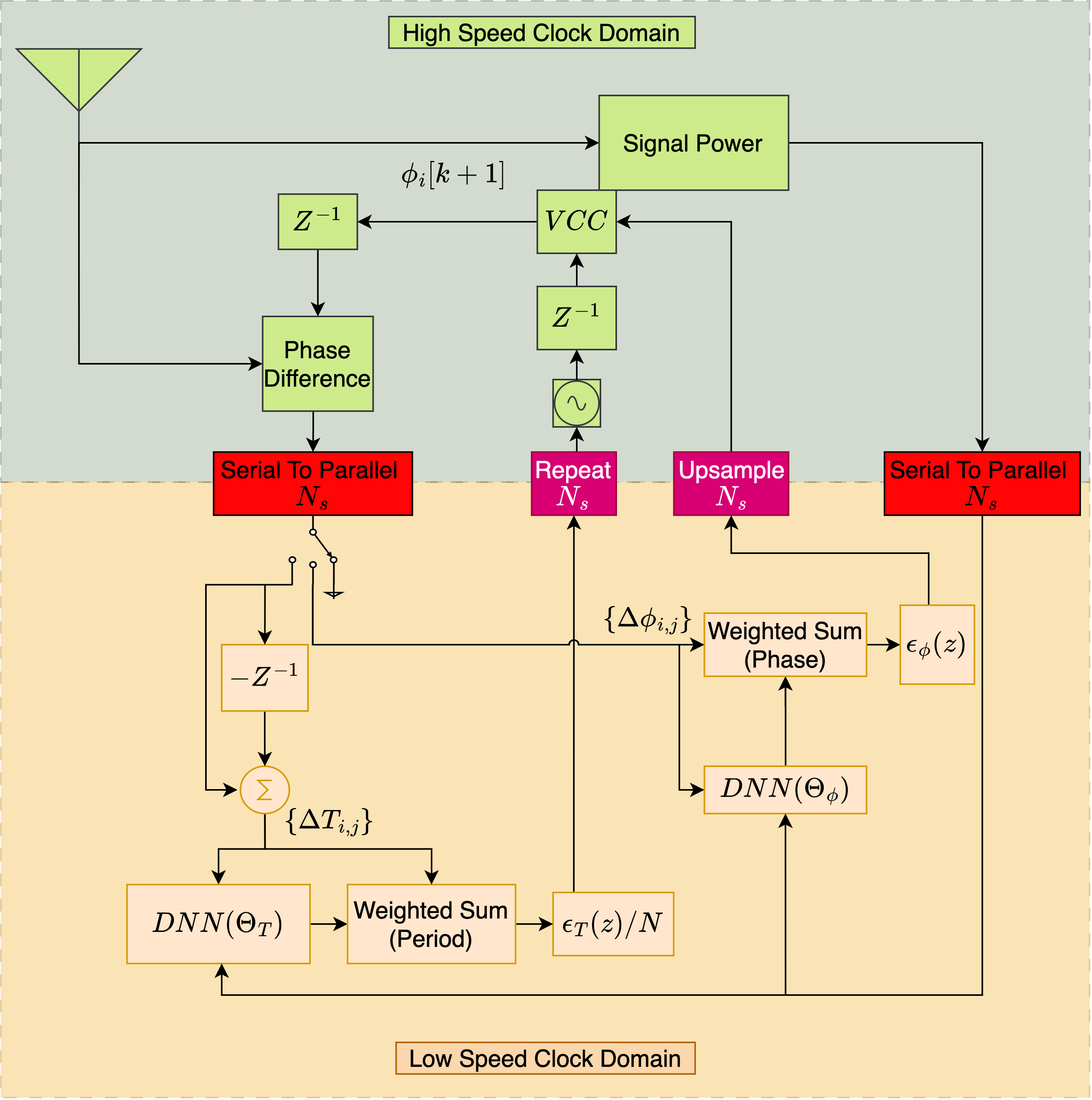}
		\caption{\small Schematic description of the proposed \ac{pfdsa}.}
		\setlength{\belowcaptionskip}{-2pt}
		\label{fig:dnn_pll_model}
	\end{figure}

In order to decouple the period update and the phase update, such that both period and phase will converge, the update takes place in a sequence of $3$ alternating actions, and therefore, a cycle of frequency and phase updates spans $3N$ times indices. We denote the interval for which $(k \mod{3N})\in\{0,2N-2\}$ as a ``collection-only" interval, at which the clocks are updated via Eqns. \eqref{eqn:Update_woCorrection} where the correction terms are set to $\Omega_i[k]=A_i[k]=0$. The ``period update" action takes place at $k$ such that $(k\mod{3N})\in\{2N-1,3N-2\}$. In this interval each node updates its period via a correction signal computed by weighting all the estimated period differences between its own period and the periods of the other nodes with weights computed by the \ac{dnn} $\psif^{(\thetaf)}(\cdot)$, which replace the weights $\alpha_{i,m}$'s used in the rule \eqref{eqn:Simeone_Extended_T}, while $\Omega_i[k]=0$. The period correction signal at node $i$ is given by
    \begin{align}
       \!\!\!\!A_i[k]
       \label{eqn:frequency_update_analyticalDNN}
       \!= \!\epsT\! \mathop{\sum}\limits_{\substack{j=1,\\ j\neq i}}^{N} \!\!\Big[\!\psif^{(\thetaf)}\!\Big(\!\Big\{\!\big(\XT^{(i)}[j'],\!  P^{(i)}[j']\big)\!\Big\}_{\substack{j'=1,\\ j'\ne i}}^N\Big) \!\Big]_j \!\!\! \cdot\! \XT^{(i)}[j].
    \end{align}
Finally, a phase correction update is applied when $(k\mod{3N})=3N-1$, by weighting the received clock phase differences with weights generated by the \ac{dnn} $\psip^{(\thetap)}(\cdot)$, which replace the weights  $\alpha_{i,m}$'s used Eqn. \eqref{eqn:Simeone_Extended_ph}, while  $A_i[k] = 0$. The phase correction signal  at node $i$ is given by
    \begin{align}
        \label{eqn:clock_analyticalDNN}
        \!\!\!\!\Omega_i[k] 
        \!=\!\epsp \mathop{\sum}\limits_{\substack{j=1,\\j\neq i}}^{N} \!\Big[\psip^{\!\!(\thetap)}\!\Big(\!\Big\{\!\!\big(\XP^{(i)}[j'],\!  P^{(i)}[j']\big)\!\!\Big\}_{\substack{j'=1,\\ j'\ne i}}^N\Big) \Big]_j \!\!\! \cdot\! \XP^{(i)}[j].
    \end{align}
\noindent  $\!\!\!\!$ The proposed algorithm is schematically depicted in Fig. \ref{fig:dnn_pll_model}.

%

%
%
    \section{The Proposed Unsupervised Distributed Online Training Scheme}
    \label{sec:Training Scheme}


    \subsection{Data Acquisition and Training Set Generation}
    Typically, \acp{dnn} are trained offline with data collected a-priori. However, as was already noted in \cite{ShlezingerDabora:2023}, achieving high accuracy clock synchronization in wireless networks requires that the training data corresponds to the actual parameters of the network deployment, i.e., the actual clock frequency differences, clock phases, propagation delays,  and network topology. This requires  collecting the training data  {\em  after deployment}.
    
    The data acquisition step takes place after startup and spans  $N_T$ \ac{tdma} frames, each consisting of $N$ transmissions, one from each node in the network. During this step, each node $i$ uses its initial, randomly generated \ac{dnn} parameters, $\thetap$ and $\thetaf$, to generate its clock times via the \ac{pfdsa},  while concurrently storing all its measured time stamps and received signal powers. For example, at time $k$, following the \ac{tdma} time-slot assignment, node $j=(k\!\mod N)+1$ transmits, and each node $i\in\mI_N\setminus j$ stores   its receive time stamp $t_{i,j}[k]$ and received powers $P_{i,j}[k]$ (if $P_{i,j}[k]<P_{th}$, then both stored parameters are set to $0$). 
    Thus, for every time $k$, node $i$ stores the pair $\mD^{(i)}[k]\triangleq\big\{ t_{i,j}[k], P_{i,j}[k]\big\}$, where $j=(k\!\mod N)+1$, $j\ne i$.
    Note that during the entire acquisition step, the parameters of both \acp{dnn} remain fixed. Due to the \ac{hd} transmission protocol, the acquisition step spans $N\cdot N_T$ time indices $k$.
    
    Based on the $N_T$  \ac{tdma} frames received during data acquisition, each node $i$ generates a training set containing $N_T$ samples, where each sample consists of $N-1$ 
    pairs, $\mD^{(i)}[k]$, containing the receive time stamp and received power. 
    \[
    \mD_i = \Big\{  t_{i,(k\!\mod\! N)+1}[k], P_{i,(k\!\mod{N})+1}[k]\Big\}_{\substack{k=0,\\ \!(k\!\mod\! N)\ne i-1}}^{N\cdot N_T-1}
    \]

    \subsection{Training Procedure}
    Once the acquisition step is completed, node $i$  processes the stored training data at sample $k\ge N$ as follows: Let $j= (k\mod{N}) +1$. Then, 
    if $P_{i,j}[k]>0$, node $i$ computes 
    \begin{eqnarray}
        \label{eqn:phase_difference_synthesis}
        \Delta \phi_{i,\nodeidx}[k]  & = & t_{i,\nodeidx}[k] - \phi_i[k]\\
        \label{eqn:time_difference_synthesis}
         T_{i, \nodeidx}[k] &=& t_{i, \nodeidx}[k] - t_{i, \nodeidx}[k-N+1].
    \end{eqnarray}
    Using the definition of $t_{i,\nodeidx}[k]$ and the update rules \eqref{eqn:Update_woCorrection} in   \eqref{eqn:time_difference_synthesis} we can express $T_{i, \nodeidx}[k]$ via the period and the phase correction signals as follows:
    \begin{align*}
         T_{i, \nodeidx}[k] & =\!\!\!\! \sum_{l=k-N+1}^k (\frac{A_j[l]}{N} + \Omega_j[l]) + N\cdot T_j[k-N+1]\\
        & =\! \frac{A_j[k\!-\!N\!\!+\!1]}{N}\cdot(\!N\!\!-\!j\!+\!1\!) \!+ \!\frac{A_j[k]}{N}\cdot(\!j\!-\!1\!) \\
        & \qquad +\! \Omega_j\!\!\left[\left\lfloor \frac{k}{N}\right\rfloor\!\! N\!\!-\!1\!\right]+N\cdot T_j[k-N+1].
    \end{align*}
    Observe that  $T_{i, \nodeidx}[k]$  represents node $i$'s estimate of the  \ac{tdma} frame duration of node $j$,  at time $k$.  Note that this estimate will have an offset at some values of $k$ due to the phase correction term, which is non-zero only at values of $k$ for which $(k\mod{3N})=3N-1$, and to the change in $A_j[k]$ within a set of $N-1$ samples at node $i$, which occurs  when $(k\mod{3N})=2N-1$. Recall also that $A_j[k]$ is non-zero only for $k$ s.t. $(k\mod{3N})\in\{2N-1,3N-2\}$. 
        
        Next, the features used for computing the outputs of the \acp{dnn} $\psiftheta(\cdot)$ and $\psiptheta(\cdot)$ are obtained in the following order
    \begin{eqnarray*}
        \label{eqn:reconstruction_of_Xphi}
        \XT^{(i)}[\nodeidx] & \leftarrow & \frac{1}{N}\big(\Delta \phi_{i,\nodeidx}[k] - \XP^{(i)}[\nodeidx]\big)\\
        \label{eqn:reconstruction_of_XT}
        \XP^{(i)}[\nodeidx] & \leftarrow &\Delta \phi_{i,\nodeidx}[k]\\
        P^{(i)}[j]  & \leftarrow  & P_{i,j}[k].
    \end{eqnarray*}
    Subsequently, the \acp{dnn} compute the weights  for the phase loop and the frequency loop for generating the error signals at node $i$,  using the respective $2(N-1)$ features at each \ac{dnn} according to the time regime, i.e., when $(k\mod{3N})=2N-1$, 
     the \ac{dnn} $\psif^{(\thetaf)}\!\Big(\!\Big\{\!\big(\XT^{(i)}[j'],\!  P^{(i)}[j']\big)\!\Big\}_{\substack{j'=1,\\ j'\ne i}}^N\Big)$ is used for computing $A_i[k]$ via \eqref{eqn:frequency_update_analyticalDNN}. Then, $A_i[k]$ is applied for period update when $(k\mod{3N})\in\{2N-1,3N-2\}$. When $(k\mod{3N})=3N-1$,  a phase correction is applied using \eqref{eqn:clock_analyticalDNN} with $\Omega_i[k]$ obtained using the \ac{dnn}  $\psip^{\!\!(\thetap)}\!\Big(\!\Big\{\!\!\big(\XP^{(i)}[j'],\!  P^{(i)}[j']\big)\!\!\Big\}_{\substack{j'=1,\\ j'\ne i}}^N\Big)$.
    Letting $ \dsone_{(\mX)}$ denote the indicator function for the event $\mX$, the loss functions are computed as follows: 
\begin{align}
        \label{eqn:loss_phi}
        \!\!\!\!\mySet{L}_{\mySet{D}_i}(\myVec{\theta}_{i, \phi}) & =\!\!\!\! \sum_{k=1}^{N_T\cdot N-1}\!\!\!\! \dsone_{(P^{(j)}[\nodeidx]>0)}\log(k+1)( t_{i, \nodeidx}[k] - \phi_i[k])^2\\
        \label{eqn:loss_T}
        \!\!\!\!\mySet{L}_{\mySet{D}_i}(\myVec{\theta}_{i, T})\! &= \!\!\!\!\!\sum_{k=1}^{N_T\cdot N-1} \!\!\!\!\!\! \dsone_{(P^{(j)}[\nodeidx]>0)}\log(k+1)\Big(\frac{T_{i, \nodeidx}[k]}{N}  \!-\! T_i[k]\Big)^2\!\!.
\end{align}

By the structure of the loss functions \eqref{eqn:loss_phi} and \eqref{eqn:loss_T}, it follows that the gradient of the loss with respect to the weights can be computed via backpropagation through time. Moreover, the loss functions can be computed in an unsupervised manner by each node locally,  facilitating unsupervised local training via conventional first-order based optimizers.
The training algorithm is summarized in Algorithm \ref{alg:UnsupLocTraining_alg}.
\RestyleAlgo{ruled}
\begin{algorithm}
    \label{alg:UnsupLocTraining_alg} 
    \KwData{Data set $\mySet{D}_i$ consisting of $N_T$ tuples, each of length $2\cdot(N-1)$; learning rate $\mu$; initial weights $(\thetap, \thetaf)$, number of epochs $E_s$ and $E_{\text{loop}}$, number of epochs for each loop $E_{\text{loop}}$, initial phases, $\{\phi_i[k]\}_{k=0}^{N-1}$.}
    \For{${\rm epoch}_1=1$ to $E_s$}{
    \textbf{Initialize:} for any $j=1, 2, ..., N$, $j\ne i$: \\
    \textbf{Read} $\left\{t_{i,j}[j-1],P_{i,j}[j-1]
    \right\}\in\mD_i$\\
        \If{ $P_{i,j}[j-1]>0$}{         
        \textbf{Set} $\XP^{(i)}[j]=t_{i,j}[j-1]-\phi_i[j-1]$; \\
        }
        \For{$m=1$ to $2\cdot E_{\text{loop}}$}{
            \For{$k'=N$ to $N\cdot N_T -1$}{
                \textbf{Set current Tx node} $\nodeidx = (k' \mod{N}) + 1$\\
                \textbf{Read} $\left\{t_{i,j}[k'],P_{i,j}[k']
                \right\}\in\mD_i$\\
                \If{ $P_{i,j}[k']>0$ \mbox{\rm and} $j\ne i$}{
                \textbf{Compute} $\Delta \phi_{i,\nodeidx}[k'] = t_{i,j}[k']-\phi_i[k']$ \\
                \textbf{Store}  $\XT^{(i)}[\nodeidx] \leftarrow (\Delta \phi_{i,\nodeidx}[k'] - \XP^{(i)}[\nodeidx])/N$ \\
                \textbf{Store} $\XP^{(i)}[j] \leftarrow \Delta \phi_{i,j}[k']$ \\
                \textbf{Store}  $P^{(i)}[j]\leftarrow P_{i,j}[k']$\\
                \textbf{Forward pass}  $P^{(i)}[j]$,  $\XT^{(i)}[\nodeidx]$ and $\XP^{(i)}[\nodeidx]$  to compute $\phi_i[k'+1]$ and $T_i[k'+1]$   via Eqns. \eqref{eqn:Update_woCorrection},  \eqref{eqn:frequency_update_analyticalDNN}, and \eqref{eqn:clock_analyticalDNN}, according to update action selected by the value of $(k\!\!\mod{3N})$
                }
            }  
            \eIf{$m \le E_{\text{loop}}$}{
                        \textbf{Compute loss} $\mySet{L}_{\mySet{D}_i}(\myVec{\theta}_{i, T})$ via Eqn. \eqref{eqn:loss_T} using the computed $T_i[k']$ and the stored $t_{i,j}[k']$'s (from which $T_{i, \nodeidx}[k']$ is computed via \eqref{eqn:time_difference_synthesis});\\
                        \textbf{Compute gradient} $\nabla_{\myVec{\theta}_i}\mySet{L}_{\mySet{D}_i}(\myVec{\theta}_{i, T})$ using backpropagation through time\;
                        \textbf{Update weights} via $\myVec{\theta}_{i, T} \leftarrow \myVec{\theta}_{i, T} - \mu \cdot \nabla_{\myVec{\theta}_{i, T}}\mySet{L}_{\mySet{D}_i}(\myVec{\theta}_{i, T})$.
            }{
                        \textbf{Compute loss} $\mySet{L}_{\mySet{D}_i}(\myVec{\theta}_{i, \phi})$ via Eqn. \eqref{eqn:loss_phi} 
                        using the computed $\phi_i[k']$ and the stored $t_{i,j}[k']$;   \\
                        \textbf{Compute gradient} $\nabla_{\myVec{\theta}_i}\mySet{L}_{\mySet{D}_i}(\myVec{\theta}_{i, \phi})$ using backpropagation through time\;
                        \textbf{Update weights} via $\myVec{\theta}_{i, \phi} \leftarrow \myVec{\theta}_{i, \phi} - \mu \cdot \nabla_{\myVec{\theta}_{i, \phi}}\mySet{L}_{\mySet{D}_i}(\myVec{\theta}_{i, \phi})$.
            }

        } 

    }
    \caption{ Unsupervised~Online Local Training at Node $i$}
\end{algorithm}
Observe that at each training batch, first, only the frequency loop is  trained, and then only the phase loop is trained. This is done to avoid the phase loop compensating for frequency errors, which would decrease the accuracy of  frequency synchronization across the nodes.

We note that as training is local at each node, based only on its received inputs with fixed \acp{dnn} parameters during training data acquisition, then the proposed approach is  flexible and scales with the size of the network without requiring any modifications. 


\subsection{ Discussion and Complexity Analysis}
The \acp{dnn} used in the frequency synchronization loop and in the phase synchronization loop have the same structure, each consisting of a total of six layers, including two linear layers, each followed by a sigmoid layer, and a third linear layer followed by a softmax layer, which is the output of the \ac{dnn}. The output of the \ac{dnn} is then passed to a selector which outputs only the weights corresponding to nodes whose signal was received above the reception threshold. These weights are then normalized such that their sum is one and are then applied in the loop. 
The number of inputs and outputs of each layer  are summarized in Table \ref{tab:DNN_Inputs_and_Outputs}.

\begin{table}[ht]
\centering
\small
    \begin{tabular}{ |c|c|c|c| } 
    \hline
    Operation     & Inputs   & Outputs \\
    \hline
    Linear Layer  1 & $2(N-1)$ & $30$  \\ 
    Sigmoid Layer 1 & $30$     & $30$  \\ 
    Linear  Layer 2 & $30$     & $30$  \\  
    Sigmoid Layer 2 & $30$     & $30$  \\
    Linear  Layer 3 & $30$     & $N-1$ \\
    Softmax Layer  & $N-1$    & $N-1$ \\
    Selection      & $N-1$    & $N-1$ \\
    Normalization  & $N-1$    & $N-1$ \\
    \hline
    \end{tabular}
        \vspace{0.4cm}
    \caption{\small Summary of the layers in the DNN implementation.}
    \label{tab:DNN_Inputs_and_Outputs}
\end{table}

For a network with $N$ nodes, each  \ac{dnn} consists of $(3(N-1)+30)\cdot 30$ weights and $2\cdot 30 + N-1$ biases. For a network with $N=16$ nodes, we obtain $2250$ weights and $75$ biases, which corresponds to less than $2.5\cdot10^3$ products for inference per \ac{dnn}. 
We note that each \ac{dnn} is used for inference only once per $3\cdot N$ time slots, thus the rate of computation at a node is less than $1700$ products for inference per \ac{tdma} frame. As discussed in \cite{ShlezingerDabora:2023}, such a computational burden  is  feasible on real-time  modern micro-controllers.

\section{Performance Evaluation}
\label{sec:simulations}

In the simulations we used the baseline setup described in Sec. \ref{subsec:baseline}. In this section we  shall depict the comparison only with the ESSBS algorithm described in Sec. \ref{subsec:stateoftheart}, 
as the  algorithms based on skew and offset estimation, e.g., \cite{maggs2012consensus}, 
achieve worse performance than ESSBS for the scenarios tested in this section. In the learning phase of the simulations, the \ac{pfdsa} algorithm was trained using  $E_s=6$ cycles with $E_{\text{loop}}=5$. Thus, overall each \ac{dnn} is trained over $30$ epochs with a learning rate of $\mu=0.1$. The information for training was collected over $N_T=126$ \ac{tdma} frames. As each \ac{tdma} frame facilitates transmission of $N=16$ nodes, then the overall time samples collected for training at a given node is $2016$. In the testing phase, $N_T=751$ \ac{tdma} frames were used in the evaluation.

\begin{figure}
    \centering
    \includegraphics[width=0.95\linewidth]{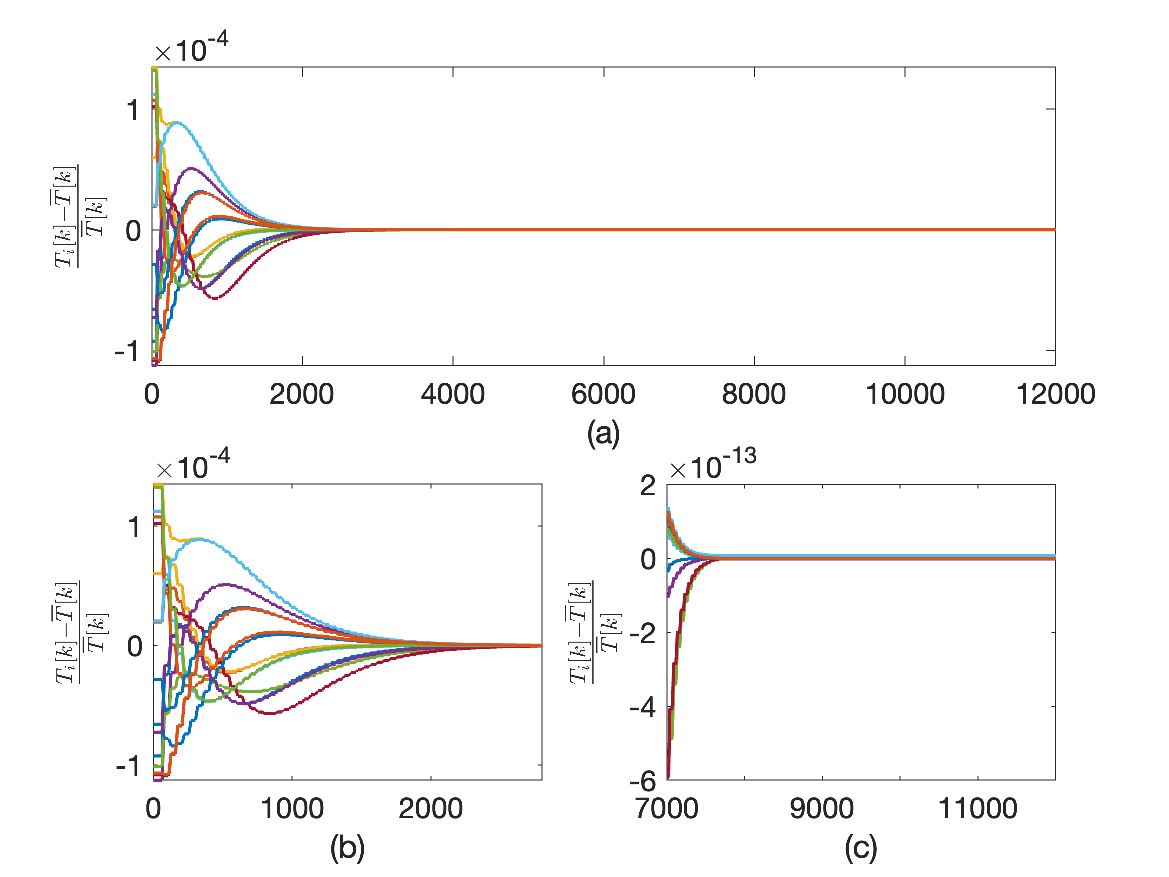}
    \caption{\small The evolution of the clock periods $T_i[k]$ vs. $k$, for the \ac{pfdsa} for all nodes (each line represents the period at a different node in the network), $\varepsilon_T=\varepsilon_{\phi}=0.3$.}
    \label{fig:Period_vs_k_eps_0p3}
\end{figure}

\begin{figure}
    \centering
    \includegraphics[width=0.95\linewidth]{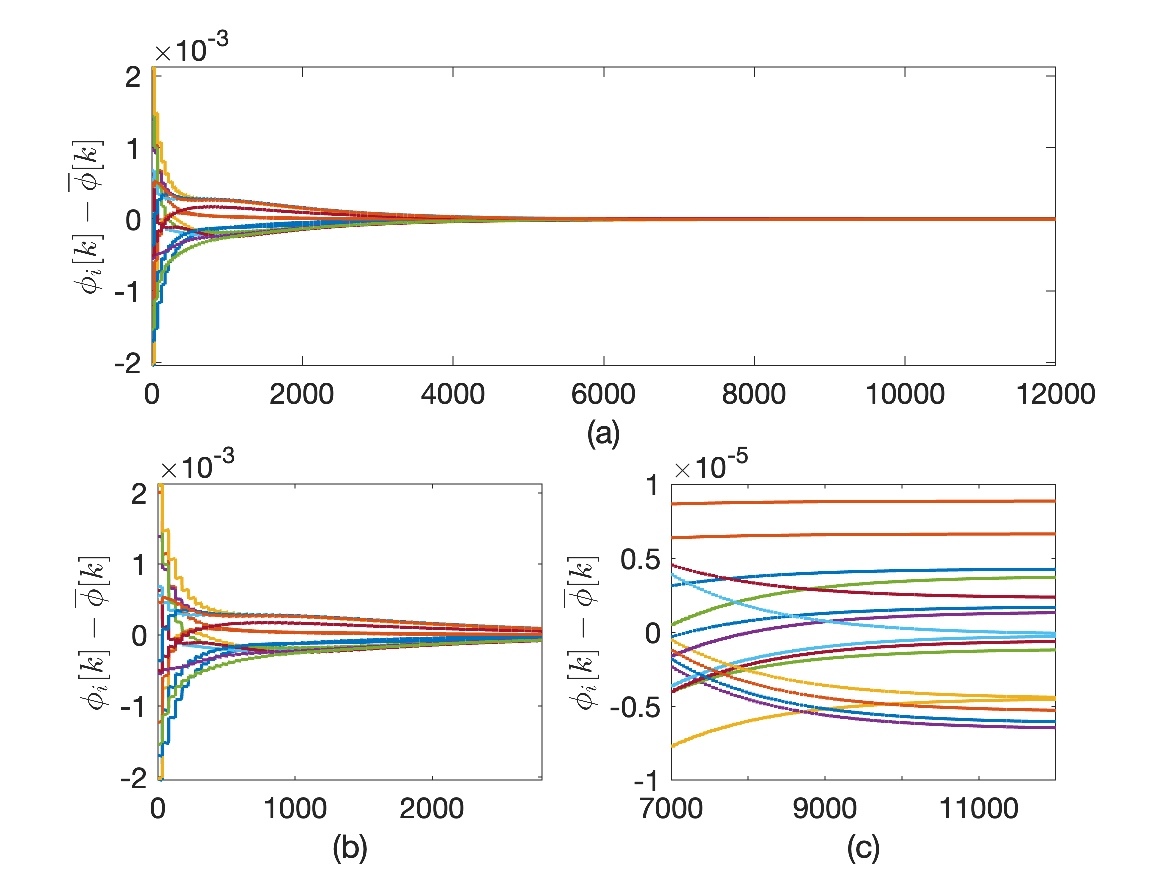}
    \caption{\small The evolution of the clock phases $\phi_i[k]$ vs. $k$, for the \ac{pfdsa} for all nodes (each line represents the phase at a different node in the network), $\varepsilon_T=\varepsilon_{\phi}=0.3$.}
    \label{fig:Phases_vs_k_eps_0p3}
\end{figure}

\begin{figure}
    \centering
    \includegraphics[width=0.83\linewidth]{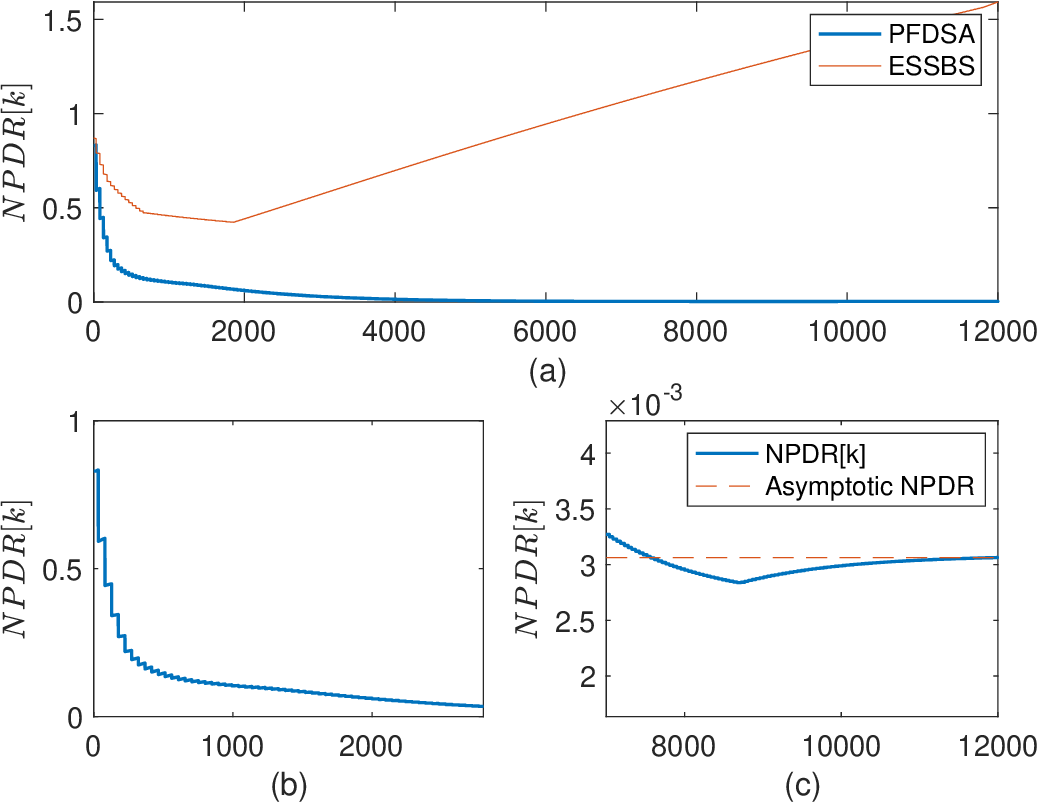}
    \caption{\small The evolution of the \ac{npdr} vs. $k$, (a) for  \ac{pfdsa} and ESSBS algorithm, $\varepsilon_T=\varepsilon_{\phi}=0.3$.; subfigs. (b) and (c) show regions of interest only for the PFDSA.}
    \label{fig:NPDR_vs_k_eps_0p3}
\end{figure}

The results for the proposed \ac{pfdsa} algorithm for the network depicted in Fig. \ref{fig:Geographical_Locations} are presented in Figs. \ref{fig:Period_vs_k_eps_0p3} (synchronization of the period) and \ref{fig:Phases_vs_k_eps_0p3} (synchronization of the phase). Both figures present the quantities w.r.t. the respective instantaneous mean. The figures demonstrate that the proposed \ac{pfdsa} algorithm is indeed able to achieve excellent frequency and phase synchronization performance, which is considerably better than that of the ESSBS algorithm. The main performance characteristics, namely, the \ac{npdr}, is depicted in Figure \ref{fig:NPDR_vs_k_eps_0p3}a. It is observed that the proposed \ac{pfdsa} is able to achieve accurate synchronization while ESSBS is not able to synchronize the phase at all.


\begin{figure}
    \centering
    \includegraphics[width=1.0\linewidth]{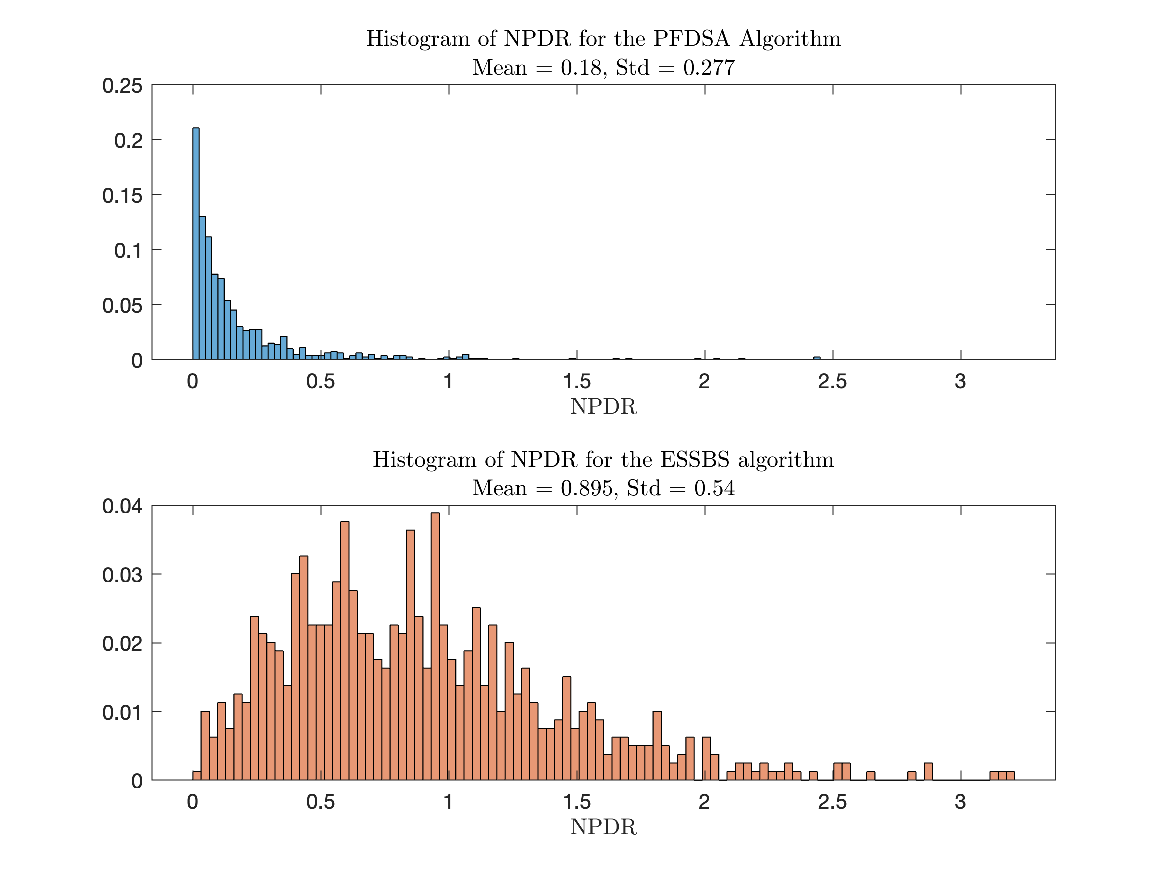}
    \caption{\small The histogram of the \ac{npdr} for \ac{pfdsa} (top) and for ESSBS (bottom), $\varepsilon_T=\varepsilon_{\phi}=0.3$.}
    \label{fig:Hist_NPDR}
\end{figure}

\begin{figure}
    \centering
    \includegraphics[width=1.0\linewidth]{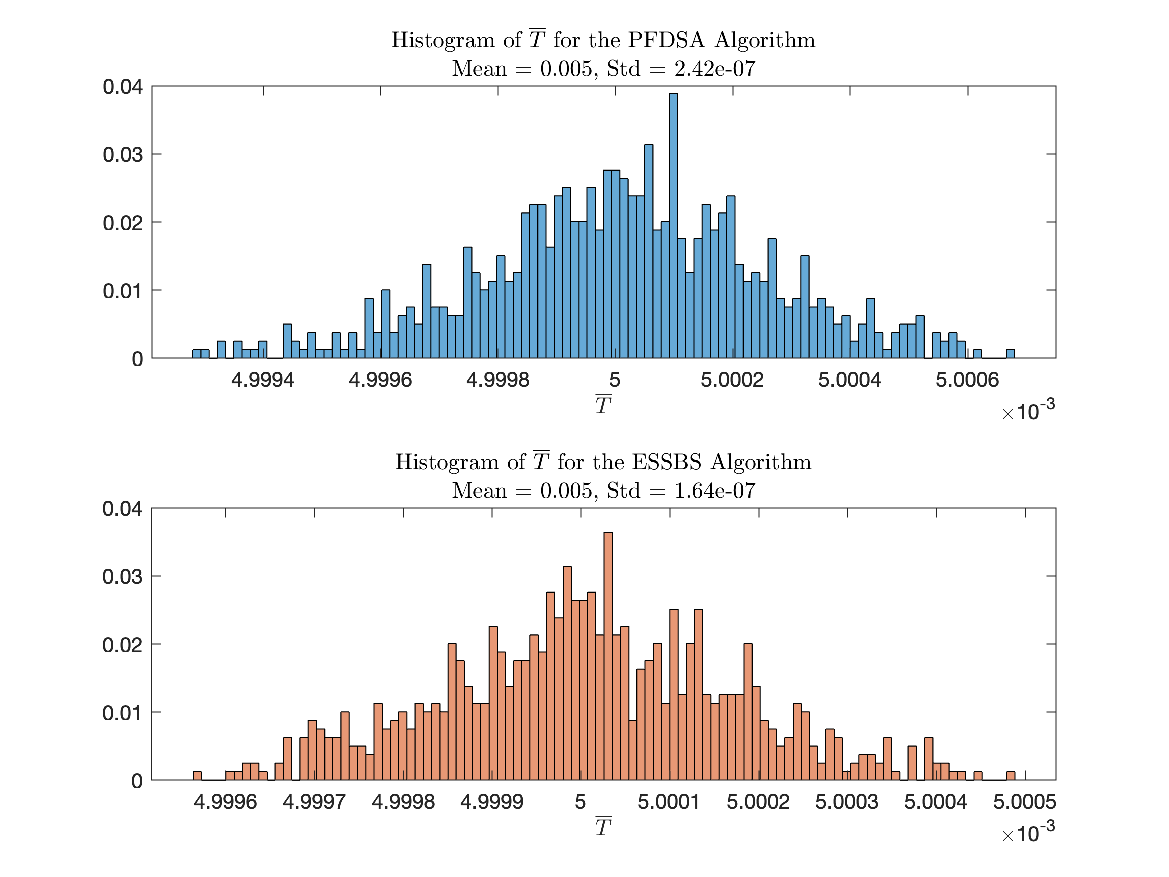}
    \caption{\small The histogram of the mean period for \ac{pfdsa} (top) and for ESSBS (bottom), $\varepsilon_T=\varepsilon_{\phi}=0.3$.}
    \label{fig:Hist_T}
\end{figure}

Next, we tested the impact of network parameters on the performance, by randomly generating  $800$ network scenarios in which the nodes were randomly placed in the network area, and the initial phases and periods were selected according to the statistics described in Sec. \ref{subsec:baseline}.  Only scenarios in which about 30\% of the links are received above the detection threshold were considered.
The histogram of the \ac{npdr} and of the mean period are depicted in Figs. \ref{fig:Hist_NPDR} and \ref{fig:Hist_T}, respectively. It is observed from the figure that both algorithms can typically be expected to achieve accurate frequency synchronization. Yet, the ESSBS would typically result in very high \ac{npdr} which implies that it is not able to achieve phase synchronization, and therefore does not facilitate communications over the network. In contrast, the proposed \ac{pfdsa} would typically achieve very accurate phase synchronization and represents a major improvement both in the mean \ac{npdr} (a factor of $4.5$) and in its \ac{std} (a factor of $2$).

\color{black}

	\section{Conclusions}
	\label{sec:conclusion}

In this work we have considered the problem of clock synchronization in networks operating in the \ac{hd} regime. 
Synchronization of both clock phase and clock frequency introduces several challenges: The first challenge is that phase correction can compensate for frequency differences, and thus, we proposed a scheme in which these two updates are decoupled. The second challenge is the stability and convergence of the synchronization scheme.  In this work we use a nested loop structure, where each loop is a first order loop. This structure benefits from the  stability and convergence associated with such loops. The third challenge follows as facilitating accurate synchronization requires that the synchronization parameters correspond to the scenario parameters, however, to date, the optimal discriminator for a given scenario parameters is unknown. In particular, attempting to apply the classic weights of \cite{simeone2008distributed} results in very poor synchronization performance. We address this challenge by employing \ac{dnn}-based phase and period discriminators for the \ac{hd} regime. Combined with a temporal update regime that facilitates  decoupling of the phase and the frequency updates, the proposed approach achieves significant improvement compared to the (generalized) classic approach of \cite{simeone2008distributed}. 
Future work will focus on further improving the  performance of the \ac{dnn}-based scheme by considering different initialization procedures, and evaluation of performance for different channel conditions relevant to wireless communications.


\begin{spacing}{0.99}
\bibliographystyle{IEEEtran}
\bibliography{IEEEabrv, myreference.bib}

\begin{thebibliography}{1}
\providecommand{\url}[1]{#1}
\csname url@samestyle\endcsname
\providecommand{\newblock}{\relax}
\providecommand{\bibinfo}[2]{#2}
\providecommand{\BIBentrySTDinterwordspacing}{\spaceskip=0pt\relax}
\providecommand{\BIBentryALTinterwordstretchfactor}{4}
\providecommand{\BIBentryALTinterwordspacing}{\spaceskip=\fontdimen2\font plus
\BIBentryALTinterwordstretchfactor\fontdimen3\font minus \fontdimen4\font\relax}
\providecommand{\BIBforeignlanguage}[2]{{%
\expandafter\ifx\csname l@#1\endcsname\relax
\typeout{** WARNING: IEEEtran.bst: No hyphenation pattern has been}%
\typeout{** loaded for the language `#1'. Using the pattern for}%
\typeout{** the default language instead.}%
\else
\language=\csname l@#1\endcsname
\fi
#2}}
\providecommand{\BIBdecl}{\relax}
\BIBdecl

\bibitem{simeone2008distributed}
O.~Simeone, U.~Spagnolini, Y.~Bar-Ness, and S.~H. Strogatz, ``Distributed synchronization in wireless networks,'' \emph{IEEE Signal Process. Mag.}, vol.~25, no.~5, pp. 81--97, Sep. 2008.

\bibitem{ShlezingerDabora:2023}
E.~Abakasanga, N.~Shlezinger, and R.~Dabora, ``Unsupervised deep-learning for distributed clock synchronization in wireless networks,'' \emph{IEEE Trans. Veh. Technol.}, vol.~72, no.~9, pp. 12\,234--12\,247, 2023.

\bibitem{simeone2007distributed}
O.~Simeone and U.~Spagnolini, ``Distributed time synchronization in wireless sensor networks with coupled discrete-time oscillators,'' \emph{EURASIP J Wirel. Commun. Netw.}, vol. 2007, pp. 1--13, Jun. 2007.

\bibitem{AerialNetworks:2020}
W.~Jaafar, S.~Naser, S.~Muhaidat, P.~C. Sofotasios, and H.~Yanikomeroglu, ``Multiple access in aerial networks: From orthogonal and non-orthogonal to rate-splitting,'' \emph{IEEE Open J. Veh. Technol.}, vol.~1, pp. 372--392, 2020.

\bibitem{TDMAnavigation:2020}
J.~Zhu and S.~S. Kia, ``A spin-based dynamic {TDMA} communication for a {UWB}-based infrastructure-free cooperative navigation,'' \emph{IEEE Sens. Lett.}, vol.~4, no.~7, pp. 1--4, 2020.

\bibitem{TDMASenser:2020}
J.~C. López-Ardao, R.~F. Rodríguez-Rubio, A.~Suárez-González, M.~Rodríguez-Pérez, and M.~E. Sousa-Vieira, ``Current trends on green wireless sensor networks,'' \emph{Sensors}, vol.~21, no.~13, 2021.

\bibitem{maggs2012consensus}
M.~K. Maggs, S.~G. O'Keefe, and D.~V. Thiel, ``Consensus clock synchronization for wireless sensor networks,'' \emph{IEEE Sens. J.}, vol.~12, no.~6, pp. 2269--2277, Jun. 2012.

\bibitem{Zino:Inpreparation}
I.~Zino, R.~Dabora, and H.~V. Poor, ``Unsupervised distributed learning for accurate clock synchronization in half-duplex {TDMA} networks,'' \emph{{\em in preparation}}.

\end{thebibliography}
	\IEEEaftertitletext{\vspace{-100\baselineskip}}
\end{spacing}

\end{document}